\documentclass[a4paper]{article}

\usepackage{18lomcon}        
\usepackage{cite}             
\usepackage{epsfig}           
\usepackage{epstopdf}
\usepackage{amssymb}

\begin{document}


\title{THE COHERENT WEAK CHARGE OF MATTER}

\author{Alejandro Segarra \email{alejandro.segarra@uv.es}
}

\affiliation{Departament de F\'isica Te\`orica and IFIC, Universitat de Val\`encia
-- CSIC,\\ C/Dr. Moliner 50, E-46100 Burjassot (Spain)}


\date{}
\maketitle


\begin{abstract}
We study the long-range force arising between two aggregates of
ordinary matter due to a neutrino-pair exchange, in the limit of zero
neutrino mass. Even if matter is neutral of electric charge, it is
charged for this weak force. The interaction is described in terms of
a coherent charge, which we call the weak flavor charge of aggregated
matter. For each one of the interacting aggregates, this charge
depends on the neutrino flavor as $Q_W^{\nu_e} = 2Z-N$,
$Q_W^{\nu_\mu} = Q_W^{\nu_\tau} = -N$, where $Z$ is the number of protons
and $N$ the number of neutrons. $Q_W^{\nu_e}$ depends explicitly on $Z$
because of the charged current contribution to $\nu_e e$ elastic
scattering, while the $N$ term in the three charges comes from the
universal neutral current contribution. The effective potential
describing this force is repulsive and decreases as $r^{-5}$. Due to
its specific behavior on $(Z,N)$ and $r$, this interaction is
distinguishable from both gravitation and residual electromagnetic
forces. As neutrinos are massive and mixed, this potential is valid
for $r\lesssim 1/m_\nu$, distances beyond which a Yukawa-like attenuation kicks
in.
\end{abstract}

\section{Introduction}
%

The study of the origin of neutrino mass is one of the
directions in which we can expect finding new Physics, even though its
small value ($m_\nu \lesssim 1$ eV \cite{taup, PDG}) makes it hard to
observe experimentally. As well as determining the absolute mass of
the neutrino, there's still a more fundamental question about their
nature unanswered: 
their finite mass could be explained through a Dirac mass term
(implying there is a conserved total lepton number $L$ distinguishing neutrinos from antineutrinos,
which are described by $4-$component Dirac spinors) or through a Majorana one
(implying that neutrinos are self-conjugate of all charges, described
by $2$ independent degrees of freedom).

In any case, the fact that their masses are very low stands, and we
discuss here another property of neutrinos as mediators of a new
force. As is well known, the processes represented in Quantum Field
Theory by the exchange of a massless particle give raise to long-range
interactions. An easy example is the scattering of two particles
mediated by a photon, which---at tree level---describes Coulomb
scattering. Our objective in this work is the application of these
ideas to a process mediated by neutrinos. According to the Electroweak
Lagrangian, the lowest-order process is a neutrino-pair exchange,
which---since neutrinos are nearly massless---describes an interaction
of long range.

\section{Long-range weak interaction between aggregate matter}

We are interested in obtaining the interaction potential due to a neutrino-pair
exchange between two matter aggregates, say $A$ and $B$. In doing so, we will
not impose any restriction on the internal structure of the
aggregates---whatever they are, we only ask them to be neutral of electric
charge. Therefore, for each aggregate, its composition is specified by two
numbers: $Z$ will represent the number of protons and the number of electrons,
which must be the same, and $N$ will represent the number of neutrons.

\begin{figure}[b!]
	\begin{minipage}{0.4\textwidth}
		\centering
		\includegraphics[height=3.9cm]{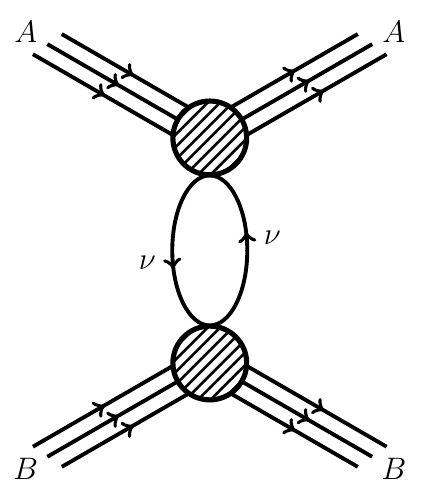}\hspace{2pc}
		\caption{\label{fig:ABAB} Effective neutrino-pair exchange interaction
			between two aggregates of matter. The blobs in the vertices
		represent any structure the aggregates could have.}
	\end{minipage} 
	\hfill
	\begin{minipage}{0.5\textwidth}
		\centering
		\includegraphics[height=3.5cm]{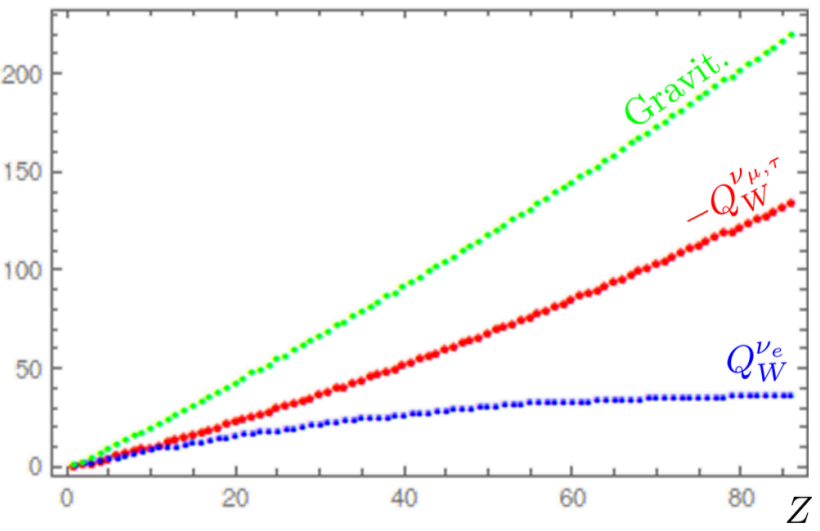}\hspace{2pc}
		\caption{\label{fig:charges} Weak flavor
				charges of the elements specified by the atomic number $Z$,
				compared to their mass $\approx Z+N$. The isotope chosen is the
				one in which the $(Z,N)$ pair lies in the valley of nuclear
			stability.}
	\end{minipage}\hfill
\end{figure}

The picture is now clear. Elastic
interactions of matter constituents with neutrinos is through either $W$ or $Z$
exchange, as well as the aggregate structure,
determine the 1-loop neutrino-pair exchange $AB\to AB$ elastic interaction, as
shown in Fig.\ref{fig:ABAB}.

The whole interaction potential between the two aggregates is given by the
Fourier Transform of the Feynman amplitude in Fig.\ref{fig:ABAB}. Since we are
only interested in the long-range part of the potential, a few simplifications
can be performed. Through rewriting the amplitude as an unsubtracted dispersion
relation, we find the long-range behavior is fully determined by its
absorptive part. In turn, the absorptive part is determined, after
unitarity-cutting the diagram in the $t$-channel, by a simple tree-level
$A\nu\to A\nu$ amplitude. This tree-level calculation is straightforward---in
the process, we only kept the dominant contributions, neglecting both
incoherent and relativistic corrections.

The analysis described above leads \cite{TFM} to the interaction potential
\begin{equation}
	V(r)= \frac{G_F^2}{8\pi^3}\,\left[ (2Z_A-N_A)(2Z_B-N_B) + 2 N_A N_B \right]\, \frac{1}{r^5} \,.
	\label{eq:V}
\end{equation}

\section{Coherent weak flavor charges}
A carefoul reading of Eq.(\ref{eq:V}) shows the standard structure of an
interaction potential. By defining, for each of the aggregates, their weak
flavor charges as $Q_W^{\nu_e} = 2Z-N$ and $Q_W^{\nu_\mu} = Q_W^{\nu_\tau} =
-N$, one gets the usual structure $V = (\mathrm{coupling})^2\times (\mathrm{product
of charges})\times (\mathrm{power law})$.

Indeed, this shows that, within the Standard Model, matter is charged! The
values of the weak charges of all stable atoms are represented in
Fig.\ref{fig:charges}. At this point, we remark that the unique $(Z,N)$
dependence of these charges makes them scale with the size of the system in a
different way than gravitation. Also, the fact that the charges of all elements
have the same sign (for each flavor) implies that this interaction is always
repulsive. These two properties may become crucial in disentangling this weak
interaction from gravitation experimentally.

\section{Prospects}
The long-range potential obtained in this work, Eq.(\ref{eq:V}), is
valid and of interest for distances between nanometers and
microns. The short-distance limit comes from the requirement of having
neutral (of electric charge) systems of aggregate matter, while the
long-distance limit is imposed by a non-vanishing value of the
absolute mass of the neutrino---indeed, the range of this interaction
for neutrinos of $m\sim 0.1$ eV is of the order of $R \sim 1/m_\nu \sim 1\mu$m.
In this region, the effective potential will become of Yukawa type instead of
the pure inverse power law.

The neutrino mass dependence of the effective potential in the
long-range behavior opens novel directions in the study of the most
interesting pending questions on neutrino properties: absolute
neutrino mass (from the range), flavor dependence and mixing (from the
weak charges in the interaction) and, hopefully, with two neutrino
exchange, the exploration of the most crucial open problem in neutrino
physics: whether neutrinos are Dirac or Majorana particles.

\section*{Acknowledgments}
The author acknowledges the Spanish MECD financial support through the FPU14/04678 grant.


\end{document}